\documentclass[10pt]{article}

\usepackage{fullpage}
\usepackage{setspace}
\usepackage{parskip}
\usepackage{titlesec}
\usepackage{placeins}
\usepackage{xcolor}
\usepackage{lineno}
\usepackage{amsmath}
\usepackage[numbers,sort&compress]{natbib}

\PassOptionsToPackage{hyphens}{url}
\usepackage[colorlinks = true,
            linkcolor = blue,
            urlcolor  = blue,
            citecolor = blue,
            anchorcolor = blue]{hyperref}
\usepackage{etoolbox}

\renewenvironment{abstract}
  {{\bfseries\noindent{\abstractname}\par\nobreak}\footnotesize}
  {\bigskip}

\titlespacing{\section}{0pt}{*3}{*1}
\titlespacing{\subsection}{0pt}{*2}{*0.5}
\titlespacing{\subsubsection}{0pt}{*1.5}{0pt}

\usepackage{authblk}

\usepackage{graphicx}
\usepackage[space]{grffile}
\usepackage{latexsym}
\usepackage{textcomp}
\usepackage{longtable}
\usepackage{tabulary}
\usepackage{booktabs,array,multirow}
\usepackage{amsfonts,amsmath,amssymb}
\providecommand\citet{\cite}
\providecommand\citep{\cite}

\newif\iflatexml\latexmlfalse
\AtBeginDocument{\DeclareGraphicsExtensions{.pdf,.PDF,.eps,.EPS,.png,.PNG,.tif,.TIF,.jpg,.JPG,.jpeg,.JPEG}}

\usepackage[utf8]{inputenc}
\usepackage[english]{babel}

\usepackage{float}

\begin{document}

\title{Molecular Frame Photoelectron Angular Distributions in Polyatomic Molecules from Lab Frame Coherent Rotational Wavepacket Evolution}

\author[1]{Margaret Gregory}%
\author[2]{Paul Hockett}%
\author[2,3,4,5]{Albert Stolow}%
\author[1]{Varun Makhija\thanks{vmakhija@umw.edu}}
\affil[1]{Department of Chemistry and Physics, University of Mary Washington, 1301 College Avenue, Fredericksburg VA, 22401.}%
\affil[2]{National Research Council of Canada, 100 Sussex Drive, Ottawa, ON, K1A 0R6, Canada}%
\affil[3]{Department of Physics, University of Ottawa, 150 Louis Pasteur, Ottawa, ON, K1N 6N5, Canada}%
\affil[4]{Department of Chemistry and Biomolecular Sciences, University of Ottawa, 150 Louis Pasteur, Ottawa, ON, K1N 6N5, Canada}
\affil[5]{Joint Centre for Extreme Photonics, NRC-University of Ottawa, ON, K1A 0R6, Canada}

\vspace{-1em}

  \date{\today}

\begingroup
\let\center\flushleft
\let\endcenter\endflushleft
\maketitle
\endgroup

\begin{abstract}
    A theory and nethod for a matrix-based reconstruction of Molecular Frame (MF) photoelectron angular distributions (MFPADs) from Laboratory Frame (LF) measurements (LFPADs) is developed and basic applications are explored. As with prior studies of MF reconstruction, the experimental side of this protocol is based upon time-resolved LF measurements, in which a rotational wavepacket is prepared and probed as a function of time via photoionization, followed by a numerical reconstruction routine. In contrast to other methodologies, the protocol presented here does not require determination of the photoionization matrix elements, and consequently takes a relatively simple numerical form of a matrix equation. Significantly, this simplicity allows the successful reconstruction of MFPADs for polyatomic molecules. We numerically demostrate this scheme for two realistic molecular photoionization cases: $N_2$ and $C_2H_4$. This new technique is expected to be generally applicable to a broad range of MF reconstruction problems involving photoionization of polyatomic molecules. 
\end{abstract}

\section{Introduction}

Photoelectron Angular Distributions (PADs) from isolated molecules provide a wealth of information about the state of the molecule from which the electron is ejected. In the case of ionization of a stationary, ground state molecule the outgoing electron experiences the molecular potential which imprints an orbital angular momentum dependent phase shift on its scattering wavefunction. The relative phases of the various angular momentum channels (or partial waves) which comprise the scattering wavefunction determine the shape of the PAD. One may therefore infer both electronic and vibrational structural information from the PAD~\cite{cohen1966,dill1976,cooper1969,reid2003,Reid2012,williams2012,kushawaha2013}. This is equally true of excited molecular electronic states, whose properties may be directly inferred from PADs~~\cite{Seideman2001,neumark2001,stolow2004,underwood2008}. The information content of the measurement is greatly enhanced if Molecular Frame PADs (MFPADs) can be measured. In such cases, the partial wave content of the scattering wavefunction is greatly incresed, since there is no orientational averaging, thus making available additional information on the molecular and electronic structure~\cite{shigemasa1995,reid2003,underwood2008,Reid2012,Yagishita2015}. For example, if MFPADs can be measured for a number of laser polarization directions relative to the molecular axes, a quantum tomography of the final state (the ejected electron plus the ion) is possible in some cases by retrieval of the relative phase of each partial wave channel. This is traditionally termed a `complete experiment'~~\cite{gessner2002,lucchese2002,hockett2009,reid1992,leahy1991,lebech2003,tang2010,marceau2017}. Recent experiments and related studies have also demonstrated direct measurement of molecular structure from such an experiment~\cite{williams2012,rescigno2012,douguet2013,kushawaha2013,menssen2016}. These features of MFPADs may also be applied to the case of a time-evolving excited state - i.e. molecular wavepacket dynamics. In such cases both MFPADs and LFPADs can track the time varying electronic character of the excited state, thus allowing measurement of time-resolved electronic-vibrational dynamics, including their non-adiabatic coupling~\cite{reid2003,underwood2008,bisgaard2009time,wu2011,rescigno2012, douguet2013, makhija2020}. MFPADs, as in the static case, provide a greater depth of information and potentially directly imprint the electronic character of the wavepacket onto the photoelectron observable~\cite{bisgaard2009time, underwood2008}. For a general review of the fundamentals of PADs see~\cite{hockett2018QMP1,hockett2018QMP2}.

It is therefore of considerable interest to develop methods that will allow the measurement of MFPADs in non-linear polyatomic molecules, the primary goal of this manuscript. There are several experimental approaches which can be used to determine MFPADs. Conceptually these rely on either (i) transforming to the MF via vector correlations in the LF, or (ii) alignling molecular axes in the LF to some degree. The first is based on kinematically complete (6D) determination of the 3D recoil momentum vectors of both photoelectrons and associated ionic photofragments in coincidence~\cite{lucchese2002,golovin1997,ito2000,gessner2002,menssen2016, shigemasa1995,heiser1997,takahashi2000,downie1999}. A fundamental restriction of this approach is that it applies only to dissociative photoionization \footnote{Or requires a post-ionization dissociation laser pulse to achieve this, as used in Coulomb explosion imaging (CEI) type  measurements.}. Nevertheless, this approach was used to study ultrafast nonadiabatic photodissociation of a polyatomic molecule~\cite{gessner2006}. Combined with the axial recoil approximation, this permits transformation of the LFPAD into the recoil frame which is often very close to that of the MF. The axial recoil approximation, however, is very restrictive in that it requires fragments to be emitted axially along the bond direction, such as in favourable cases containing carbon-halogen bonds. Unfortunately, molecules in excited states generally bend as they dissociate, thus obviating broader use of the axial recoil approximation~\cite{lam2020}. Importantly, the general approach presented here depends on neither dissociative photoionization processes nor the axial recoil approximation.

The second approach to MFPADs is via laser-aligning or orienting molecular ensembles prior to photoionization, for which numerous experimental techniques have been developed, and are thoroughly reviewed in~\cite{koch2019} and~\cite{stapelfeldt2003}. Briefly, ensembles of linear molecules with a high degree of laser-induced alignment are readily achievable under field-free conditions~\cite{cryan2009,ren2013,marceau2017}, while orientation of molecules lacking inversion symmetry is much harder to achieve~\cite{fleischer2011,egodapitiya2014,de2009,damari2016,oda2010,kitano2013}. In addition, the experimentally achievable degree of ground state alignment directly impacts the ability to extract MFPADs~\cite{reid2018}. For non-linear polyatomic molecules, aligning or orienting all three body-fixed (MF) axes under field-free conditions has also achieved for some cases, but still presents a significant hurdle to MF measurements in general~\cite{lin2018,ren2014,lee2006,underwood2005}. Even in cases where all axes can be orientated, an arbitrary molecular orientation cannot be accessed since all rotational degrees of freedom cannot be independently controlled. Finally, for the extension of these techniques to excited state vibrionic wavepacket studies, one must keep in mind that the coherent MF axis distributions constructed for the molecular ground state are modified in the excited state due to the pump transition dipole matrix elements which themselves have polarization sensitive selection rules. In general, one cannot expect that the excited state axis distribution moments, following pump pulse absorption, will be the same as those for the MF ground state. This presents a further challenge to the use of selected rotational revivals for the `fixed-in-space’ approach to MFPADs experiments.

To overcome these hurdles, an alternative route to the MF is available via time domain Rotational Wavepacket (RWP) measurements, specifically the Orientation Resolution through Rotational Coherence Spectroscopy (ORRCS) method~\cite{lam2020,sandor2019,sandor2018,marceau2017,makhija2016,mikosch2013,mikosch22013,ramakrishna2012}. In this approach, a RWP is launched by an impulsive nonresonant femtosecond laser pulse (the pump pulse) in a thermal molecular gas. Rather than trying to simultaneously fix all three axes in space at a specific time delay, this method generates a coherent ground state rotational wavepacket (RWP). The RWP is then photoionized by a delayed femtosecond probe pulse, and the signal of interest is measured as a function of the time delay. This time dependent signal provides sufficient information in some cases to permit reconstruction of the signal for a perfectly oriented molecule, at an arbitrary orientation with respect to the LF. The ORRCS approach has been demonstrated for the cases of strong field ionization of linear ~\cite{sandor2019} and non-linear polyatomic molecules~\cite{makhija2016,mikosch2013,sandor2018}, strong field photodissociation of linear molecules~\cite{lam2020}, molecular high harmonic generation spectroscopy~\cite{ren2013,wang2017,wang2020}, and photoionization matrix element retrieval (hence MFPAD reconstruction) for one-photon ionization of a homonuclear diatomic molecule.~\cite{marceau2017}. 

Here we show that the ORRCS approach applies to the reconstruction of MFPADs for the photoionization of polyatomic molecules, without any requirement for dissociative photoionization, or the more stringent requirements of complete photoionization matrix element retrieval (as recently demonstrated for a homonuclear diatomic in Ref.~\cite{marceau2017}). A summary of the general approach we develop is as follows. We first reduce the problem to a system of linear equations
\begin{equation}
    \mathbf{C}^{mol} = \mathbf{\hat{G}} \mathbf{C}^{lab},
    \label{eq:basic}
\end{equation}
by applying symmetry considerations. Here $\mathbf{C}^{mol}$ and $\mathbf{C}^{lab}$ are vectors in the molecular and lab frames respectively, and $\mathbf{\hat{G}}$ is a known matrix. Components of the vector $\mathbf{C}^{lab}$ can be directly extracted from time resolved measurements of LFPADs from one-photon ionization of an evolving RWP. Provided that a sufficient number of components are extracted, we can uniquely determine $\mathbf{C}^{mol}$ by solving Eq.~\ref{eq:basic}, which is directly related to the MFPADs for various polarization geometries. The structure of matrix $\mathbf{\hat{G}}$ determines the number of $\mathbf{C}^{mol}$ that can be retrieved, which then determines for which polarization geometries the MFPAD can be reconstructed. We detail this route to the MFPADs, as follows. In section~\ref{sec:Theory}, we outline the theory that leads to an equation like Eq.~\ref{eq:basic}. This is further divided into subsections. In section~\ref{sec:Rot-dynamics}, we briefly discuss the formalism used to describe the ground state RWP. This leads to section~\ref{sec:AF-ionization}, in which we discuss the formalism of photoionization from a RWP. In section~\ref{sec:MF-ionization} we descirbe the related formalism for ionization in the MF. This then leads to section~\ref{sec:AF-MF-theory}, in which we relate photoinzation in the MF and LF and invoke symmetry srguments to derive the desired Eq.~\ref{eq:CfromG} - an equation of the form of Eq.~\ref{eq:basic} - for inversion symmetric molecules having both a horizontal and vertical mirror plane. In section~\ref{sec:MF-recon}, we demonstrate the reconstruction of MFPADs, for various polarizations, for photoionization of $N_2$ ($D_{\infty h}$ point group symmetry) and the asymmetric top $C_2 H_4$ ($D_{2h}$ point group symmetry) to their ground cationic electronic states. Finally, in section~\ref{sec:Conclusion} we provide a summary of our work, potential future avenues of investigation and suggested experimental applications. 

\section{Theory\label{sec:Theory}}
In the method proposed here MFPADs, are reconstructed using LFPADs measured during the time evolution of a RWP. First, a non-resonant, femtosecond duration, linearly polarized intense pump pulse intersects a molecular gas, exciting via stimulated Raman Scattering several rotational states in the ground electronic state~\cite{stapelfeldt2003, koch2019}. Next, a time delayed probe pulse with photon energy greater than the molecular ionization potential ionizes the molecule and the momentum distribution of the ejected photoelectrons is measured. This measurement is repeated at several time delays, resulting in a time sequence of LFPADs. These LFPADs serve as raw material for the reconstruction described here. We begin discussing the theoretical formalism needed to describe this experiment, based on the formalism developed by Underwood and coworkers~\cite{underwood2000,underwood2005,underwood2008}. In the following, we first discuss the description of the rotational wavepacket (Sec. \ref{sec:Rot-dynamics}), followed by the description of the ionization at each time delay (Sec. \ref{sec:AF-ionization}). In the subsequent subsection (Sec. \ref{sec:MF-ionization}) we discuss photoionization in the MF, and finally derive the connection between the LFPADs and MFPADs that allow us to go from the experiment to the MFPAD (Sec. \ref{sec:AF-MF-theory}), which is the main result of this work. For further background to the current work, interested readers can find more detailed introductory material on the topic of ionization from aligned molecular ensembles in, for example, Refs. \cite{Seideman2001,underwood2008,hockett2018QMP1}, and references therein. Readers interested purely in the applications of the protocol may wish to skip the derivations and theoretical discussion, and proceed directly to Sec.~\ref{sec:MF-recon}, which illustrates, numerically, MF reconstruction demonstrated here for two cases.

The overarching aim of this section is to reduce the LF-MF connection to a solvable system of linear equations which of the form of Eq.~\ref{eq:basic}. Here $\mathbf{C}^{mol}$ and $\mathbf{C}^{lab}$ are vectors in the molecular and lab frames respectively, and $\mathbf{\hat{G}}$ is a known matrix. Components of the vector $\mathbf{C}^{lab}$ can be directly extracted from time resolved measurements of LFPADs from one-photon ionization of an evolving RWP. Provided that a sufficient number of components are extracted, one can uniquely determine $\mathbf{C}^{mol}$ by solving Eq.~\ref{eq:basic}, should such a solution prove to exist.

\subsection{Ionization from the Rotational Wavepacket}

\subsubsection{Describing the Rotational Dynamics\label{sec:Rot-dynamics}}

In our formalism, the evolution of the excited rotational wavepacket can be described by the time and orientation angle dependent molecular axis distribution~\cite{underwood2000,underwood2005,underwood2008},
\begin{equation}
    P(t,\phi,\theta,\chi) = \sum_{KQS}A^K_{QS}(t)D^{K*}_{QS}(\phi,\theta,\chi).
    \label{eq:MAD}
\end{equation}
This equation is a multipole expansion of the probability - $P(t,\phi,\theta,\chi)$ - that, at a time $t$ after the excitation, the molecule will have an orientation described by the Euler angles $(\phi, \theta, \chi)$. These angles serve to define the orientation of a rigid body in three dimensions~\cite{goldstein}, with respect to a space-fixed (LF) coordinate system. The space-fixed coordinate system is labelled by $(X,Y,Z)$, and a body-fixed (MF) coordinate system by $(x,y,z)$, in the standard manner~\cite{zare1988}. In the space-fixed frame the $Z$-axis is defined by the laser polarization and the $Y$-axis by the laser propagation direction. The molecular axes define the molecular frame $(x,y,z)$, and are chosen according to molecular geometry, with $z$ usually corresponding to the axis of highest symmetry. The Wigner D matrix elements $D^K_{QS}(\phi,\theta,\chi)$ are a basis on the space of orientations~\cite{sakurai} and facilitate the expansion. It may be noted that, were the third Euler angle $\chi$ set to zero (as would be the case for linear molecules), the Wigner D matrix elements reduce to the well known spherical harmonics~\cite{zare1988}. In the more general case of a 3D rigid rotor, the third angle is required to describe the orientation of the rotor about $z$-axis. 

Note that the time dependence of the wavepacket is fully described by the functions $A^K_{QS}(t)$. These multipole moments of the axis distribution, or axis distributions moments (ADMs), are exactly analogous to the spherical multipole moments of any electrostatic charge distribution~\cite{jackson}, with the fundamental difference being the inclusion of an additional index needed due to the dependence on a third angle - note the three expansion indices $K, Q$ and $S$, rather than the usual indices $l$ and $m$ of the (speherical harmonic) multipole expansion. 
The index $K$ takes integer values and corresponds to $l$ (degree or rank of the operator, often corresponding physically to total angular momentum) in the standard multipole expansion; the remaining indicies define components (order of the operator, or projection terms) within the ($K$ or $l$) subspace. For instance, $K = 1$ is a dipole moment, and if the coefficient $A^1_{00}$ dominates the expansion at a particular time, then the axis distribution strongly resembles $\cos\theta$ at that time and so on. The indices $Q$ and $S$ each run from $-K$ to $K$, by analogy with $m$ which runs from $-l$ to $l$, and describe the axis distribution as a function of the angles $\phi$ and $\chi$ respectively. Also analagous to $l$, $K$ admits the physical interpretation of angular momenta associated with the rotational wavepacket. The maximum value of $K$ can be thought of as the maximum angular momentum transferred to the molecule by the pump\footnote{Strictly speaking, this is the maximum angular momentum present in the axis distribution of the LF ensemble, defined to include all rotational state contributions, see Ref. \cite{underwood2008} for details.} and, analogous to $m$, $Q$ is its projection on the LF $Z$ axis. The new index $S$ is the projection of $K$ on the molecular $z$ axis, and is conjugate to the angle $\chi$ that tracks the rotation of the molecular body around the $z$ axis.          

Finally we note the restrictions on the values of $K$, $Q$ and $S$ which are imposed by symmetry. For a wavepacket excited by a linearly polarized field, the molecular axis distribution must be independent of $\phi$ in order to preserve cylindrical symmetry, thus rendering $Q = 0$. Further, $K$ must be even to preserve inversion symmetry.  Although the equations herein are general, we usually restrict discussion to this special (but common) case, and drop the additional subscript where it is unnecessary. (This inversion and cylindrically-symmetric case is generally termed \textit{molecular alignment}, and is conceptually distinct from \textit{molecular orientation}, which also allows non-cylindrically symmetric cases with up-down asymmetry, hence additionally yielding odd $K$ and $Q\neq 0$ terms.) The restrictions on $S$ depend on the point group symmetry of the molecule, and are dealt with individually for the specific cases analyzed here (Sec. \ref{sec:MF-recon}). 

\subsubsection{Describing the Ionization Step\label{sec:AF-ionization}}

Excitation of a RWP by a pump pulse is followed by ionization by a time delayed probe pulse. The measured, time resolved LFPADs are also typically described by a multipole expansion as follows,
\begin{equation}
\sigma(\epsilon,t,\theta_{e},\phi_{e})=\sum_{LM}\beta_{LM}(\epsilon,t)Y^{L}_{M}(\theta_{e},\phi_{e}),
\label{LFPAD}
\end{equation}
where $\epsilon$ is the kinetic energy of the electron and $\theta_{e}$ and $\phi_{e}$ its polar and azimuthal ejection angles. The anisotropy parameters $\beta_{LM}(\epsilon,t)$ are now the time dependent multipole moments of the LFPADs, which fully characterize their variation as a result of the evolving rotational wavepacket. These therefore must contain the ADMs described above, and $L$ and $M$ must be related to $K$ and $S$ (recall that $Q=0$). The relation between the ADMs and the anisotropy for perturbative one-photon ionization was derived under the dipole approximation in~\cite{underwood2008}. The resulting expression may be compactly expressed as,
\begin{align}
\beta_{LM}(\epsilon,t)=\sum_{KS}C^{LM}_{KS}(\epsilon) A^{K}_{0S}(t)
\label{eq:betaLF}\\
C^{LM}_{KS}(\epsilon)=\sum_{\zeta\zeta'}D_{\zeta}(\epsilon)D^{*}_{\zeta'}(\epsilon)\Gamma^{\zeta\zeta'LM}_{K0S}.\label{eq:coeffLF}
\end{align}
where we have explicitly set $Q=0$ (i.e. cylindrical symmetry). 
The daunting number of subscripts and superscripts in these equations (resulting from several tensor products of irreducible representations of $SO(3)$~\cite{underwood2008}) primarily represent the angular momenta involved in the problem. We first note that, much like in Eq.~\ref{eq:MAD}, the time dependence of the $\beta_{LM}(\epsilon,t)$ is entirely contained in the ADMs as expected. The energy dependent coefficients $C_{KS}^{LM}(\epsilon)$ couple $K$ and $S$ to the angular momentum of the ejected electron $L$ and its projection $M$ on the molecular $z$ axis. The coupling is facilitated by a coupling parameter we designate $\Gamma^{\zeta\zeta'LM}_{KQS}$, which is summed over possible irreducible components of the electric field tensor - essentially arising as the square of the ionizing photon's angular momentum~\cite{underwood2008}, thus defining an effective angular momentum coupling into the observables - $P = 0, 1$ or $2$ and their projections on the molecular axis, $\Delta q = q' - q$, where $q$ and $q'$ are spherical components of the electronic dipole moment operator as discussed below. The full form of $\Gamma^{\zeta\zeta'}_{KQS}$, built from a number of Wigner 3j Symbols, is provided in Eq.~\ref{eq:FullLFGamma} in the appendix. Physically, it is of note that the terms arise from the square of the photoionization matrix elements, and include various couplings which ultimately contribute to any experimental observables; these can be treated as effective angular momentum couplings of the derived quatities. For example, the coupling between the electric field tensor rank $P$ and the ADMs with rank $K$ (hence the axis distribution $P(t,\phi,\theta,\chi)$, see Eq.~\ref{eq:MAD}), results in a term with allowed angular momentum values ranging from $L = K-P$ to $L = K + P$ and molecular frame projections $S + \Delta q$. This term therefore provides selection rules for $L$, and sets limits on the observable spatial anisotropy in the LF, $L_{max}$, which can be no higher than that defined by the anistoropy of the electric field tensor coupled to the molecular axis distribution (additional selection rules may pertain, see below and Eq.~\ref{eq:FullLFGamma}).

The remaining label $\zeta$ pertains to basis or channel functions used to represent the photoionization process in the formalism in~\cite{underwood2008}. $\zeta$ labels consist of three parts: $\zeta^+=\nu_\alpha^+$, $\alpha^+$ labels vibrational and electronic states of the ionic core, $\zeta_{dip}=q=-1$,$0$ or $1$, the spherical component of the electronic dipole moment operator which facilitates ionization, and $\zeta_f$=$\varGamma$,$\mu$,$h$,$l$ labels the basis functions used to construct the wave function of the ionized electron. This wavefunction, defined exclusively in the MF, appears in expressions for both the LF and MFPADs, and thus is the key to linking the two. In order to interpret the label $\zeta_f$, we need to analyze the form of this wavefunction, which is written as,
\begin{eqnarray}
\phi(\mathbf{r'_e},\epsilon;\mathbf{R})=\sum_{\varGamma\mu h l}X^{\varGamma\mu*}_{hl}(\theta'_e,\phi'_e)\psi_{\varGamma\mu h l}(r'_e,\epsilon;\mathbf{R}),\\
X^{\varGamma\mu}_{hl}(\theta'_e,\phi'_e)=\sum_{\lambda}b^{\varGamma\mu}_{hl\lambda}Y_{l\lambda}(\theta'_e,\phi'_e).
\label{eq:wavefunc}
\end{eqnarray} 
$\mathbf{R}$ represents the nuclear coordinates, $r'_e,\theta'_e,\phi'_e$ are coordinates of the electron in the MF, and $\epsilon$ its energy. Further, $X^{\varGamma\mu}_{hl}(\theta'_e,\phi'_e)$ are symmetry adapted spherical harmonics - a superposition of spherical harmonic functions - $Y_{l\lambda}(\theta'_e,\phi'_e)$ - which form a basis for the representation $\varGamma$ of the molecular point group. $\mu$ and $h$ are corresponding symmetry labels which replace the projection of $l$ on the $z$-axis $\lambda$ when cylindrical symmetry is lost, while accounting for any remaining degeneracy. For a linear molecule, $\lambda$ is a conserved quantity and these labels vanish. Given this equation, we can interpret $l$ as the orbital angular momentum of the ejected electron in the molecular frame. This is, however, not a good quantum number in the region of the nascent ion core (i.e., not a conserved quantity), since the molecular potential experienced by the outgoing electron is not spherically symmetric. Physically, angular momentum can be exchanged with the potential during ionization/electron scattering. Furthermore, it is also the case that the initial bound state orbital is usually described by a superposition of many $l$ terms. It is of particular note that this results, in general, in more complex scattering patterns than would pertain in the simplest case: the final state is not restricted by a $\Delta l=\pm1$ selection rule as might be intuitively expected for a simple model of single electron-photon coupling under spherical symmetry, as is often applicable in atomic photoionization problems from states with well-defined initial orbital angular momentum (see, for instance, Refs. \cite{Fano1972, reid2003, hockett2018QMP1} for further discussion). In the observable, the mixing of pairs of channels labeled by $l$ and $l'$, leads to the lab frame angular momentum $L$ with values $|l-l'|, |l-l'+1|,...,|l+l'-1|,|l+l'|$. The imperative function performed by $\Gamma^{\zeta\zeta'LM}_{K0S}$ is to simultaneously couple $K$, $S$, $l$ and $l'$ into a single lab frame angular momentum $L$. It performs the same function with respect to the relevant projections of each of these angular momenta. This connection between the scattering channels and the LF angular momentum will be imperative in the following section in which a connection between the LF and MFPADs is derived. 

With this, we can finally identify the quantities $D_\zeta(\epsilon)$ in Eq.~\ref{eq:coeffLF} as the matrix elements of irreducible components of the MF electronic dipole operator, between the ground electronic state and a continuum channel function labeled by $\zeta$. We note that it is the set of products $D_{\zeta}(\epsilon)D_{\zeta'}(\epsilon)^{*}$ that appear in the equation. We designate these as $d_{\zeta\zeta'}$, and rewrite Eq.~\ref{eq:coeffLF} as,
\begin{equation}
    C^{LM}_{KS}(\epsilon) = \sum_{\zeta\zeta'} d_{\zeta\zeta'}(\epsilon)\Gamma^{\zeta\zeta'LM}_{K0S}.
    \label{eq:coeffLF2}
\end{equation}
As shown in the next section, these coefficients $C^{LM}_{KS}$ represent the coefficients $C^{lab}$ in Eq.~\ref{eq:basic}, which can be extracted by measurements of time varying LFPADs from a rotational wavepacket. The ADMs in Eq.~\ref{eq:betaLF} can be accurately simulated for the electronic ground in the rigid rotor approximation, provided the experimental conditions are well known~\cite{mikosch22013,makhija2012,marceau2017}. Therefore, the coefficients $C^{LM}_{KS}$ can be extracted \emph{uniquely} from a measurement of the $\beta_{LM}(\epsilon,t)$ by linear regression under appropriate experimental conditions. This procedure, constituting the method of ORRCS, has been demonstrated and analyzed by several other authors for different molecules and experimental conditions, in several contexts including one-photon ionization~\cite{lam2020,sandor2019,sandor2018,marceau2017,makhija2016,mikosch2013,mikosch22013}. Therefore, for the remainder of the discussion of the reconstruction protocol developed herein, we will start with the assumption that the $C^{LM}_{KS}$ have been reliably extracted from an experimental measurement. As detailed in the previous publications, the number of coefficients extractable depends on the number of rotational states excited by the pump pulse, since a larger spread of coherently excited $J$ states enhances the contribution of the higher $K$ ADMs ~\cite{underwood2008,marceau2017}. In the following we identify the molecular frame coefficients $C^{mol}$, and derive the equivalent of Eq.~\ref{eq:basic} which will allow reconstruction of MFPADs using the extracted coefficients $C^{LM}_{KS}$.  

\subsection{Ionization in the Molecular Frame\label{sec:MF-ionization}}

In the MF,  the PAD depends on energy, the MF ejection angles of the electron $\theta'_{e}$, and  $\phi'_{e}$ and the relative orientation angles between the molecule and the laser polarization, $(\phi, \theta, \chi)$. The latter are the standard Euler angles as in Eq.~\ref{eq:MAD}, describing the orientation of a rigid body in the LF, and map to angles of the polarization vector in the MF. For linearly polarized light, $\theta$ and $\chi$ are the spherical polar and azimuthal angles of the polarization vector in the MF~\cite{makhija2012}. Here again we continue to follow the formalism developed in~\cite{underwood2008}. The MFPAD can be expanded in a basis set of spherical harmonics, giving the equivalent to Eq.~\ref{LFPAD} as follows,
\begin{equation}
\sigma(\epsilon,\phi, \theta, \chi,\theta'_{e},\phi'_{e})=\sum_{LM}\beta_{LM}(\epsilon,\phi, \theta, \chi)Y^{L}_{M}(\theta'_{e},\phi'_{e})
\label{eq.MFPAD}
\end{equation}
The MF equivalent of Eq.~\ref{eq:betaLF} and Eq.~\ref{eq:coeffLF} can be written as,
\begin{align}
\beta_{LM}(\epsilon, \phi, \theta, \chi)=\sum_{PR\Delta q}C^{LM}_{PR}(\epsilon,\Delta q) D^{P}_{R\Delta q}(\phi, \theta, \chi)    
\label{eq:betaMF}\\
C^{LM}_{PR}(\epsilon,\Delta q)=\sum_{\zeta\zeta'}d_{\zeta\zeta'}(\epsilon)\Gamma^{\zeta\zeta'LM}_{PR\Delta q}   
\label{eq:coeffMF}
\end{align}
Note that the value of $\Delta q$ is fixed in the sum over $\zeta$ and $\zeta'$ in Eq.~\ref{eq:coeffMF}, instead of taking all allowed values as in Eq.~\ref{eq:betaLF} (see Appendix for further details). \\\\
In the MF, $R$ is the projection of $P$ on the polarization axis, thus it can take the values $R= -P...P$. For linearly polarized light, $R=0$. Additionally, $\Delta q$ is the projection of $P$ on the MF axis. As in Eq.~\ref{eq:MAD}, the $D^{P}_{R\Delta q}(\phi, \theta, \chi)$ in Eq.~\ref{eq.MFPAD} are elements of the Wigner D matrix, which are analytically known irreducible representations of $SO(3)$. Thus, if the $C^{LM}_{PR}(\epsilon,\Delta q)$ can be somehow determined from the experimental measurement of $C^{LM}_{KS}$ (Eqs.~\ref{eq:coeffLF},~\ref{eq:coeffLF2}), the $\beta_{LM}(\epsilon,\phi,\theta,  \chi)$ can be determined providing the MFPAD, Eq.~\ref{eq.MFPAD}. The $D^{P}_{R\Delta q}(\phi, \theta, \chi)$, and by extension the angular momentum of the ionizing photon determines the orientation angle dependence of the MFPADs. The shape for a particular orientation, characterized by the anisotropy parameters $\beta_{LM}(\epsilon,\phi,\theta,\chi)$, is determined solely by channel mixing driven by the molecular potential experienced by the ejected electron. This channel mixing is codified in Eq.~\ref{eq:coeffMF} by both the ionization dipole product $d_{\zeta\zeta'}(\epsilon)$, which are the same as in Eq.~\ref{eq:coeffLF2} and the MF coupling parameter $\Gamma^{\zeta\zeta'LM}_{PR\Delta q}$. In this case, the allowed values are determined by $L = |l - l'|, |l - l' + 1|,...,|l + l' - 1|,|l + l'|$, by a given pair of partial wave channels $l$ and $l'$. Note that, unlike in the LF where $L$ must be additionally determined by the sum of the ionizing photon angular momentum $P$ and the rotational wavepacket angular momentum $K$, there is no additional restriction on $L$ in MF. Physically, this is simply due to the lack of any restriction on the observable spatial anisotropy in the MF, while the LF result corresponds, in essence, to the averaging of the MF result over all molecular axis alignments, as defined by the angular momentum coupling detailed in Sec. \ref{sec:AF-ionization}. As discussed in the introduction, the MFPAD directly conveys information on molecular structure and symmetry. 

The coefficients $C^{LM}_{PR}(\epsilon,\Delta q)$ in Eq.~\ref{eq:coeffMF} are the coefficients $C^{mol}$ in Eq.~\ref{eq:basic} that we would like to be able to extract from experimental data, specifically from the coefficients $C^{LM}_{KS}$ in Eq.~\ref{eq:coeffLF2}. In the following section we show that an equation like Eq.~\ref{eq:basic} exists, allowing this extraction for inversion symmetric molecules with a vertical and horizontal mirror plane.  

\subsection{From the Lab Frame to the Molecular Frame\label{sec:AF-MF-theory}}

To make a connection between the $C^{LM}_{KS}$ in the LF and the $C^{LM}_{PR}$, we must return to the experimentally relevant equation, Eq.~\ref{eq:betaLF}, for the measureable $\beta_{LM}(\epsilon,t)$, from which the coefficients $C^{LM}_{KS}$ can be extracted. These are then subject to Eq.~\ref{eq:coeffLF2} which we restate here,
\begin{equation}
C^{LM}_{KS}(\epsilon) = \sum_{\zeta\zeta'} d_{\zeta\zeta'}(\epsilon)\Gamma^{\zeta\zeta'LM}_{K0S}.
\end{equation}
This equation relates the $C^{LM}_{KS}$ to products of ionization transition dipole moments $d_{\zeta\zeta'}$ from the ground electronic state into a pair of scattering channels $\zeta$ and $\zeta'$ . We note that these are complex in general, and to emphasize this we write the $d_{\zeta \zeta'}$ in terms of their amplitudes and phases as $d_{\zeta \zeta'}=|d_{\zeta \zeta'}|e^{i\phi_{\zeta \zeta'}}$. Noting that $d_{\zeta' \zeta} = d^*_{\zeta \zeta'}$, Eq.~\ref{eq:coeffLF2} can be written as
\begin{equation}
C^{LM}_{KS}=\sum_{\zeta}d_{\zeta \zeta}\Gamma^{\zeta\zeta LM}_{KS} + \sum_{\zeta \zeta'<\zeta}|d_{\zeta \zeta'}|\left(\Gamma^{\zeta \zeta'LM}_{K0S}e^{i\phi_{\zeta \zeta'}} + \Gamma^{\zeta' \zeta LM}_{K0S}e^{-i\phi_{\zeta \zeta'}} \right).
\label{eq:finallfcoeff}
\end{equation}
Further, the following general relation can be derived by permutation of $\zeta$ and $\zeta'$ (cf. Appendix):
\begin{equation}
   \Gamma^{\zeta' \zeta LM}_{K0S}= (-1)^{K-S}(\Gamma^{\zeta \zeta' LM}_{K0-S})^*
   \label{eq: LFgeneral}
\end{equation}
Eq.~\ref{eq:finallfcoeff} can then be recast as,
\begin{equation}
C^{LM}_{KS}=\sum_{\zeta}d_{\zeta \zeta}\Gamma^{\zeta\zeta LM}_{K0S} + \sum_{\zeta \zeta'<\zeta}|d_{\zeta \zeta'}|\left(\Gamma^{\zeta \zeta'LM}_{K0S}e^{i\phi_{\zeta \zeta'}} + (-1)^{K-S}(\Gamma^{\zeta \zeta' LM}_{K0-S})^*e^{-i\phi_{\zeta \zeta'}} \right).
\label{eq:finallfcoeff2}
\end{equation}
Provided that a sufficient number of $C^{LM}_{KS}$ are extracted from an experimental measurement using Eq.~\ref{eq:betaLF}, one may attempt to extract several $|d_{\zeta \zeta'}|$ and their associated phases using this equation. We note, however, that in general a unique solution to this problem is not attainable, since Eq.~\ref{eq:finallfcoeff} represents a more difficult version of the restricted phase problem encountered in coherent diffractive imaging~\cite{taylor2003} or ultrafast pulse measurement~\cite{trebino2012}. In the usual restricted phase problem, the magnitudes of a set of complex numbers are directly measured and the relative phases between them are extracted by applying appropriate restrictions to the data during the extraction. Here, we cannot directly measure the magnitudes $|d_{\zeta \zeta'}|$, and must instead extract both these and their associated phases from a measurement of the $C^{LM}_{KS}$. Solving this problem constitutes a complete molecular photoionization experiment, for which a unique solution has been attained in few select cases - not including non-linear polyatomic molecules - by applying restrictions due to molecular symmetry combined with careful numerical and statistical analysis~\cite{hockett2009, reid1992,leahy1991,lebech2003, tang2010, marceau2017}. Here we instead tackle the reduced questions: What information \emph{can} we uniquely determine from a measurement of the $C^{LM}_{KS}$, and is this information sufficient to construct the MFPAD for photoionization by linearly polarized light? Note that, as discussed above, the MFPAD still directly provides information on molecular structure and symmetry since the channel mixing which determines its shape is driven by the molecular potential experienced by the ejected electron. In addition, the ionization dipole matrix elements $D_\zeta(\epsilon)$ that determine the MFPADs are themselves dependent on the nuclear configuration~\cite{hockett2009}. A means to retrieve MFPADs without the burden of a "complete" photoionization methodology - including the challenge of a full matrix element (phase) retrieval procedure~\cite{hockett2018QMP2} - is therefore desirable.      

To tackle the above question, we now turn our attention to Eq.~\ref{eq:betaLF} for the measured $\beta_{LM}(\epsilon,t)$ and investigate the connection between $d_{\zeta\zeta'}$ and experimental data. As discussed in previous articles on the rotational dynamics of asymmetric tops, see Refs.~\cite{makhija2012,pabst2010,rouzee2008,seideman1999} for details, inversion symmetry in the LF and the presence of a vertical mirror plane in the MF ensures that only ADMs of even $K$ and $S$ are excited, that the ADMs are real and that $A^K_{0S}(t) = A^K_{0-S}(t)$. Eq.~\ref{eq:betaLF} then becomes,
\begin{equation}
\beta_{L0}(\epsilon,t)=\sum_{KS=0}C^{L0}_{K0}(\epsilon)A^K_{00}(t)+\sum_{KS>0}\left(C^{L0}_{KS}(\epsilon)+C^{L0}_{K-S}(\epsilon)\right)A^K_{0S}(t).
\label{eq:D2hrotbetalf}
\end{equation}
Here we set $M = 0$ as required for cylindrical symmetry to be maintined by photoionization using linearly polarized light. By measurement of the $\beta_{L0}(\epsilon,t)$ for a fixed photoelectron energy, the $\left(C^{L0}_{KS}(\epsilon)+C^{L0}_{K-S}(\epsilon)\right)$ can be determined uniquely by linear regression if the experimental parameters are known and the $A^K_{0S}(t)$ can be accurately simulated~\cite{lam2020,makhija2016}. Using Eq.~\ref{eq: LFgeneral}, the general relationship for $\Gamma^{\zeta' \zeta LM}_{K0S}$, and Eq.~\ref{eq:finallfcoeff} for $C^{LM}_{KS}$ we get,
\begin{equation}
\frac{1}{2}\left(C^{L0}_{KS}+C^{L0}_{K-S}\right)=\sum_{\zeta}x_{\zeta \zeta}\Gamma^{\zeta \zeta L0}_{K0S}+\sum_{\zeta \zeta'<\zeta}x_{\zeta \zeta'}\left(\Gamma^{\zeta \zeta' L0}_{K0S}+\Gamma^{\zeta' \zeta L0}_{K0S}\right).
\label{eq:lffit}
\end{equation}
Now defining $\overline{C}^{L0}_{KS} = \frac{1}{2}\left(C^{L0}_{KS}+C^{L0}_{K-S}\right)$, Eq.~\ref{eq:lffit} becomes: 
\begin{equation}
\overline{C}^{L0}_{KS}=\sum_{\zeta}x_{\zeta \zeta}\Gamma^{\zeta \zeta L0}_{K0S}+\sum_{\zeta \zeta'<\zeta}x_{\zeta \zeta'}\left(\Gamma^{\zeta \zeta' L0}_{K0S}+\Gamma^{\zeta' \zeta L0}_{K0S}\right).
\label{eq:lffit2}
\end{equation}
The left hand side of this equation is a measurable number, and the right hand side is linear in the \emph{real} parameters $x_{\zeta \zeta}$ and $x_{\zeta \zeta'} = |d_{\zeta \zeta'}|\cos\phi_{\zeta\zeta'}$. 
 Thus, Eq~\ref{eq:lffit2} can be reduced to a simple matrix equation,
\begin{equation}
\mathbf{\overline{C}}^{L0}_{KS}=\mathbf{\Gamma}^{\zeta\zeta'L0}_{K0S} \mathbf{x}_{\zeta \zeta'}.
\label{eq:reducedxcoeffmfD2h}
\end{equation}
Solving this equation for $\mathbf{x}_{\zeta \zeta'}$ results in the following expression,   
\begin{equation}
    \mathbf{x}_{\zeta \zeta'}=(\mathbf{\Gamma}^{\zeta\zeta'L0}_{K0S})^+\mathbf{\overline{C}}^{L0}_{K0S} 
    \label{eq.xvalue2}
\end{equation}
Designated by $+$, we use the numerical Moore-Penrose inverse of $\mathbf{\Gamma}^{\zeta\zeta'L0}_{K0S}$, found by reduced singular value decomposition of $\mathbf{\Gamma}^{\zeta\zeta'L0}_{K0S}$~\cite{planitz1979inconsistent}.  Eq.~\ref{eq:reducedxcoeffmfD2h} is a single mathematical statement of the information content of the measurement. It tells us which pairs molecular frame scattering channels $\zeta$ and $\zeta'$ interfere to produce a LF anisotropy parameter $\beta_{L0}$. Eq.~\ref{eq.xvalue2} inverts this equation to provide the smallest vector magnitude, least-squares solution for $\mathbf{x}_{\zeta \zeta'}$ for an experimentally measured $\mathbf{\overline{C}}^{L0}_{K0S}$ vector~\cite{planitz1979inconsistent}.  We note here that this is by no means a unique inversion, since in general the number of interfering scattering channels will far outnumber the coefficients $\overline{C}^{L0}_{K0S}$. However, this does not matter provided that we can find a similar matrix equation linking $\mathbf{x}_{\zeta\zeta'}$ to the MF coefficients $C^{LM}_{PR}(\epsilon, \Delta q)$. We can then substitute for $\mathbf{x}_{\zeta\zeta'}$ using Eq.~\ref{eq.xvalue2}, eliminating $\mathbf{x}_{\zeta\zeta'}$ from the problem.

For this purpose, we now examine Eq.~\ref{eq:coeffMF},
\begin{equation}
    C^{LM}_{PR}(\epsilon,\Delta q)=\sum_{\zeta\zeta'}d_{\zeta\zeta'}(\epsilon)\Gamma^{\zeta\zeta'LM}_{PR\Delta q}   
\end{equation}
for the desired coefficient $C^{LM}_{PR}(\epsilon,\Delta q)$ in the MF. Again, we write the $d_{\zeta \zeta'}$ in terms of their amplitudes and phases as $d_{\zeta \zeta'}=|d_{\zeta \zeta'}|e^{i\phi_{\zeta \zeta'}}$. Noting that $d_{\zeta' \zeta} = d^*_{\zeta \zeta'}$, Eq.~\ref{eq:coeffMF} can be written as
\begin{equation}
C^{LM}_{PR}(\epsilon,\Delta q)=\sum_{\zeta}d_{\zeta \zeta}\Gamma^{\zeta\zeta LM}_{PR\Delta q} + \sum_{\zeta \zeta'<\zeta}|d_{\zeta \zeta'}|\left(\Gamma^{\zeta\zeta'LM}_{PR\Delta q}e^{i\phi_{\zeta \zeta'}} + \Gamma^{\zeta'\zeta LM}_{PR\Delta q}e^{-i\phi_{\zeta \zeta'}} \right).
\label{eq:finallfcoeffmf}
\end{equation}
Using a general relation, derived from the behavior of $\Gamma^{\zeta\zeta'LM}_{PR\Delta q}$ under permutation of the indices  $\zeta \zeta'$ (cf. Appendix):
\begin{equation}
    \Gamma^{\zeta' \zeta LM}_{PR\Delta q}=(-1)^{\Delta q -M
}(\Gamma^{\zeta \zeta' L-M}_{PR-\Delta q})^*
\label{eq: GeneralMF}
\end{equation}
Eq.~\ref{eq:finallfcoeffmf} can be expressed as,
\begin{equation}
C^{LM}_{PR}(\epsilon,\Delta q)=\sum_{\zeta}d_{\zeta \zeta}\Gamma^{\zeta\zeta LM}_{PR\Delta q} + \sum_{\zeta \zeta'<\zeta}|d_{\zeta \zeta'}|\left(\Gamma^{\zeta\zeta'LM}_{PR\Delta q}e^{i\phi_{\zeta \zeta'}} + (-1)^{\Delta q -M
}(\Gamma^{\zeta \zeta' L-M}_{PR-\Delta q})^*e^{-i\phi_{\zeta \zeta'}} \right).
\label{eq:finalcoeffmf}
\end{equation}
This is the MF equivalent of Eq.~\ref{eq:finallfcoeff2}. We now seek a relation similar to Eq.~\ref{eq:reducedxcoeffmfD2h}, which relates the $C^{LM}_{PR}(\epsilon,\Delta q)$ to the parameters $x_{\zeta\zeta'}$. To progress from Eq~\ref{eq:finallfcoeff2} to Eq.~\ref{eq:reducedxcoeffmfD2h}, we required that the molecule have a mirror plane. Similarly, to progress from Eq.~\ref{eq:finallfcoeffmf}, we restrict ourselves to inversion symmetric molecules with a horizontal plane. We are thus restricted to molecules with $D_{nh}$ point group symmetry. 

For such molecules, we find that (cf. Appendix)
\begin{equation}
C^{LM}_{P0}(\epsilon,\Delta q)=(-1)^{M - \Delta q}C^{L-M}_{P0}(\epsilon,-\Delta q).
\label{eq:Cequallin}
\end{equation}
This equation holds for any inversion symmetric molecule with a horizontal mirror plane, ionized by linearly polarized light (thus $R=0$). 
Using Eq.~\ref{eq:finalcoeffmf} and the identity given in Eq.~\ref{eq:Cequallin}, we add $\pm \Delta q$ terms, and simplify to get:
\begin{equation}
C^{LM}_{P0}(\epsilon,\Delta q)=\sum_{\zeta}x_{\zeta \zeta}\Gamma^{\zeta\zeta LM}_{P0\Delta q} + \sum_{\zeta \zeta'<\zeta}x_{\zeta \zeta'}(\Gamma^{\zeta\zeta'LM}_{P0\Delta q}+\Gamma^{\zeta'\zeta LM}_{P0\Delta q}) 
\label{eq:D2coeffmf}
\end{equation}
Writing Eq.~\ref{eq:D2coeffmf} as a matrix equation gives,
\begin{equation}
\mathbf{C}^{LM}_{P0}(\epsilon, \Delta q)=\mathbf{\Gamma}^{\zeta\zeta'LM}_{P0\Delta q} \mathbf{x}_{\zeta \zeta'}.
\label{eq:reducedcoeffmf2}
\end{equation}

And finally using Eq.~\ref{eq.xvalue2} for $\mathbf{x}_{\zeta\zeta'}$ in Eq.~\ref{eq:reducedcoeffmf2} gives,    
\begin{equation}
\mathbf{C}^{LM}_{P0}(\epsilon, \Delta q)=\mathbf{\hat{G}}^{LMP\Delta q }_{L'0KS} \mathbf{\overline{C}}^{L'0}_{KS} .
\label{eq:CfromG}
\end{equation}

This equation is the key result of this manuscript, of the desired form (Eq.~\ref{eq:basic}) connecting the experimentally accessible coefficients in the vector $\mathbf{\overline{C}}^{L'0}_{KS}$ to the set of coeffecients $\mathbf{C}^{LM}_{P0}(\epsilon, \Delta q)$ which can be used to construct the MFPAD (the entire reconstruction procedure is summarized at the beginning of the next section with reference to Fig.~\ref{fig:flowchart}). We denote $\mathbf{\hat{G}}^{LMP\Delta q }_{L'0KS} = \mathbf{\Gamma}^{\zeta\zeta'LM}_{P0\Delta q} (\mathbf{\Gamma}^{\zeta\zeta'L^{\prime} 0 }_{K0S})^+$, again using the Moore-Penrose inverse.  $\mathbf{\hat{G}}^{LMP\Delta q }_{L'0KS}$ is a known matrix connecting the LF to the MF for any inversion symmetric molecule containing a horizontal and vertical mirror plane, rotationally excited by a linearly polarized pump-pulse. The elements of $\mathbf{\hat{G}}^{LMP\Delta q }_{L'0KS}$ couple a particular $L'$ ($M'=0$ due to cylindrical symmetry) in the LF to a $LM$ in the MF, automatically finding sets of interfering ionization channels $\zeta$ and $\zeta'$ which contribute to both a particular $\beta_{L'0}$ in the LF and a $\beta_{LM}$ in the MF. Thus, non-zero rows in $\mathbf{\hat{G}}^{LMP\Delta q }_{L'0KS}$, corresponding to a set of $LMP\Delta q$, indicate retrievable sets of $C^{LM}_{P0}(\epsilon,\Delta q)$. If $\mathbf{\hat{G}}^{LMP\Delta q }_{L'0KS}$ is of full rank (in other words, if the set of linear equations Eq.~\ref{eq:CfromG} are consistent, or the problem is not ill-posed), then all such $C^{LM}_{P0}(\epsilon,\Delta q)$ can be uniquely determined as can, therefore, the corresponding $\beta_{LM}$ in the MF and thus the MFPADs. If not, then a least squares solution of Eq.~\ref{eq:CfromG} can be obtained~\cite{planitz1979inconsistent}, and the resulting MFPADs compared with calculated MFPADs to determine the reliability of the solution, or compared with the constraint of the experimental observables. In the next section, we numerically demonstrate the reconstruction procedure using Eq.~\ref{eq:CfromG} for the $D_{\infty h}$ molecule $N_2$ and the $D_{2h}$ molecule $C_2 H_4$.

\section{Molecular Frame Reconstruction\label{sec:MF-recon}}

\begin{figure}
\center
\includegraphics[width=0.7\columnwidth]{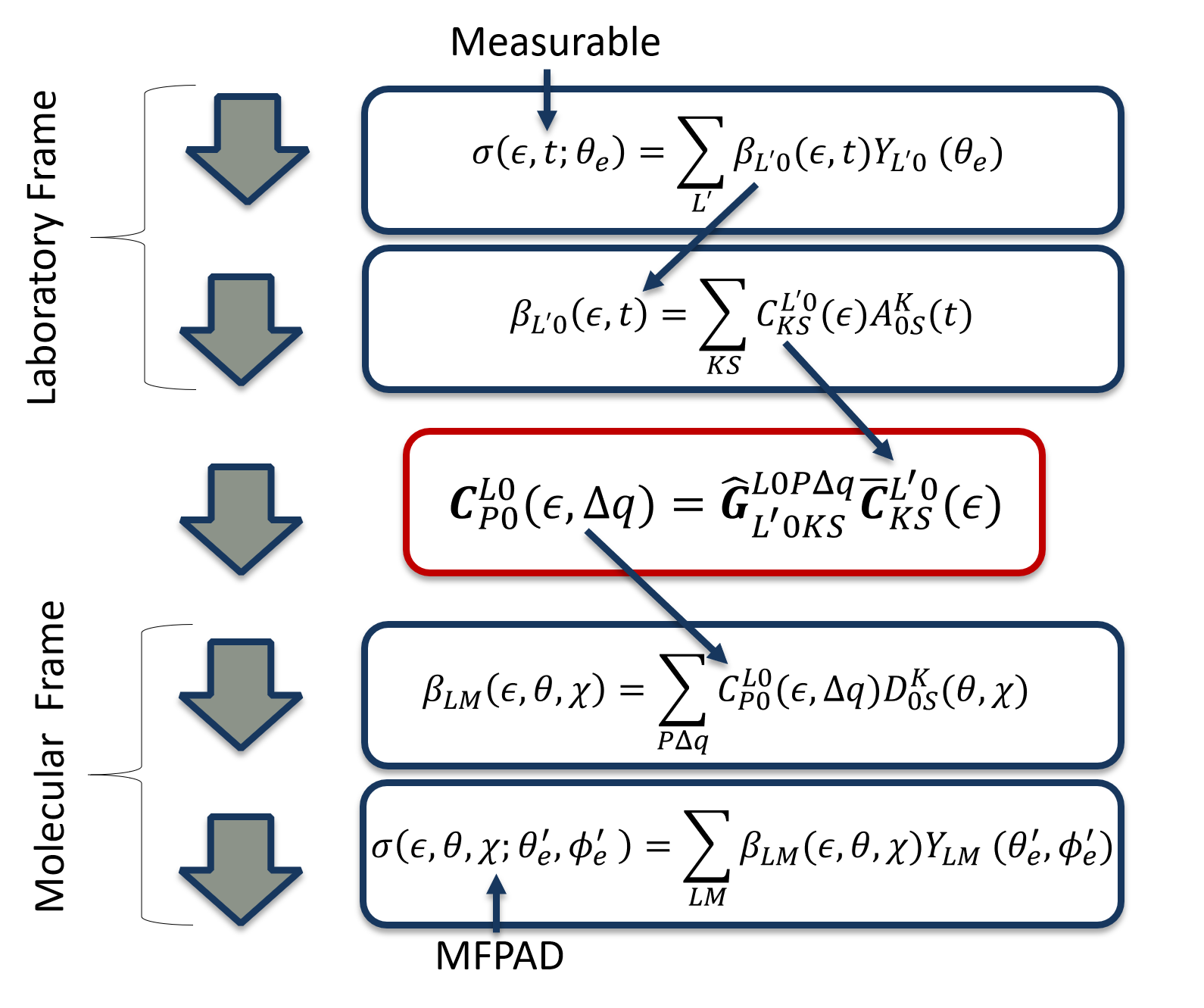}
\caption{A flow chart summarizing the MFPAD reconstruction method starting from experimental data. Measured time resolved LFPADs are the starting point, from which LF time dependent anisotropy parameters $\beta_{L'0}(\epsilon,t)$ can be determined by linear regression. Following this, the LF coefficients $C^{L'0}_{K0S}$ can be uniquely determined also by linear regression. The crucial next step, highlighted in red in the flow chart, connects the $\overline{C}^{L'0}_{KS} = 1/2(C^{L'0}_{K0S}+C^{L'0}_{K-S})$ to the MF coefficients $C^{LM}_{P0}(\epsilon, \Delta q)$ via the matrix $\mathbf{\hat{G}}^{LMP\Delta q}_{L'0KS}$. These coefficients then allow the reconstruction of the MFPADs $\sigma(\epsilon, \theta, \chi;\theta^{\prime}_e, \phi^{\prime}_e)$.}
\label{fig:flowchart}
\end{figure}

With Eq.~\ref{eq:CfromG} in hand, we can outline an MFPAD reconstruction procedure for molecules with $D_{nh}$ point group symmetry. We summarize this procedure starting from the experimentally measured, time-resolved LFPADs from an evolving rotational wavepacket excited by a linearly polarized, non-resonant pump-pulse, in Fig.~\ref{fig:flowchart}. The measured LFPAD from a evolving RWP is described by Eq.~\ref{LFPAD}, displayed in the top box of Fig.~\ref{fig:flowchart} for ionization by linearly polatized light ($M'=0$). This data can be used to extract the LF parameters $\beta_L'0(\epsilon,t)$ at a particular electron kinetic energy $\epsilon$. The time resolved anisotropy parameters $\beta_{L'0}(t)$ determined from these data can then be fit to uniquely determine the LF coefficients $C^{L'0}_{KS}$ using Eq.~\ref{eq:coeffLF}, as shown in the second box in Fig.~\ref{fig:flowchart}. This procedure has been experimentally demonstrated in previous studies~\cite{lam2020,marceau2017}. In lieu of experimental data, here we directly calculate the $C^{L'0}_{KS}$ using Eq.~\ref{eq:coeffLF2}, $C^{L'0}_{KS} = \sum_{\zeta\zeta'}d_{\zeta\zeta'}\Gamma^{\zeta\zeta'L'0}_{K0S}$. To do so, we need the product of ionization dipole matrix elements $d_{\zeta\zeta'} = D_{\zeta}D^{*}_{\zeta'}$, which we calculate using the ePolyscat ionization code~\cite{gianturco1994,natalense1999} in order to provide physically realistic values. Therefore our numerical demonstrations start with using these calculated $C^{L'0}_{KS}(\epsilon)$ in Eq.~\ref{eq:CfromG}, shown in the center red box in Fig.~\ref{fig:flowchart}. Using these we retrieve the MF coefficients $C^{LM}_{P0}(\epsilon,\Delta q)$. Then moving further down the flowchart in the figure, we can use Eq.~\ref{eq:betaMF} and Eq.~\ref{eq.MFPAD} shown in the bottom two boxes to construct the MFPADs $\sigma(\epsilon,\theta,\chi;\theta_e^{\prime},\phi_e^{\prime})$ for ionization of the molecule by linearly polarized light. 

We can also calculate the MFPADs directly from the dipole matrix elements using Eq.~\ref{eq:coeffMF}, Eq.~\ref{eq: GeneralMF} and Eq.~\ref{eq:betaMF}. We can then compare these directly calculated MFPADs to those reconstructed from the LF coefficients $C^{L'0}_{KS}$ using Eq.~\ref{eq:CfromG} in the central red box of Fig.~\ref{fig:flowchart} to test the reliability of the method. To use Eq.~\ref{eq:CfromG}, we also need to calculate the numerical matrix $\mathbf{\hat{G}}^{LMP\Delta q}_{L'0KS}$, which we can do using the analytical equations Eq.~\ref{eq:FullLFGamma} and Eq.~\ref{eq:MFFullGamma} (cf. Appendix) for the LF and MF coupling parameters $\Gamma^{\zeta\zeta'L0}_{K0S}$ and $\Gamma^{\zeta\zeta'LM}_{P0\Delta q}$. This matrix in fact determines which $C^{LM}_{P0}(\epsilon, \Delta q)$ can be determined, and in turn which MFPADs can be determined. As discussed below, in some cases $\mathbf{\hat{G}}^{LMP\Delta q}_{L'0KS}$ has rows that are entirely zero, indicating that the particular LF $L'$ and MF $L$ do not share any common pairs of photoionzation channels $\zeta$,$\zeta'$. This prevents the extraction of the corresponding $C^{LM}_{P0}(\epsilon, \Delta q)$ and therefore the corresponding $\beta_{LM}$ in the MF. This limits the MFPADs that can be extracted using this method. 

\subsection{Results for $D_{\infty h}$ Symmetry}

A molecule with $D_{\infty h}$ symmetry in the ground state constitutes a linear rigid rotor. For such a molecule, $S=0$ and we adhere to the convention $\chi=0$~\cite{zare1988}. Additionally, the MFPAD is independent of the azimuthal orientation angle $\phi$ due to cylindrical symmetry, which we set to zero. Here, we attempt to retrieve MFPADs for ionization of $N_2$ from its neutral ground electronic state, producing photoelectrons with $\epsilon=$7.6~eV by one-photon ionization. This choice is based on previous work~\cite{marceau2017} in which a complete experiment was performed for the same continuum channels at this photoelectron energy, using a RWP in the ground state. 

\begin{figure}
    \centering
    \includegraphics[width=.9\textwidth,trim=15 15 15 15, clip]{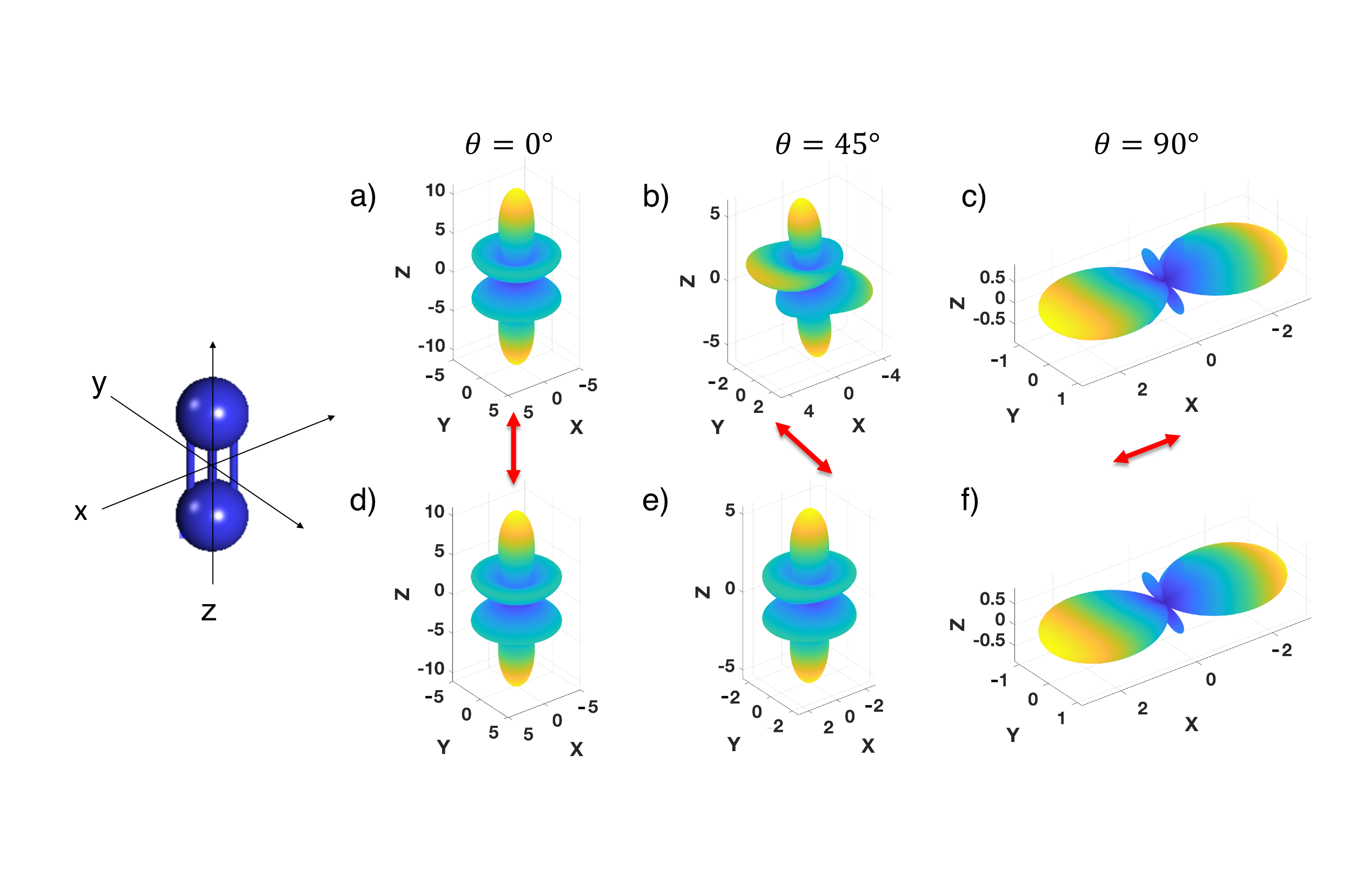}
    \caption{Calculated ((a),(b),(c)) and retrieved ((d),(e),(f)) MFPADs for $N_2$ with light polarized (red arrows) along the $z$ axis, at $\theta = 45^\circ$ with respect to the $z$ axes, and along the $x$ axis respectively. The MF is shown to the lef, with $x$, $y$ and $z$ axes corresponding to those in the MFPAD plots.}
    \label{fig:MFPADs}
\end{figure}

 To numerically demonstrate our methodology, we first calculated the LF coefficients $C^{L'0}_{K0}(\epsilon)$ from values of the ionization dipole matrix elements - $D_{\zeta}(\epsilon)$ - calculated using ePolyscat code at $\epsilon = 7.6$~eV \cite{gianturco1994,natalense1999}. We provide these values for each ionization channel in Table~\ref{tab:Dinfinitytable} along with quantum numbers that specify each channel, and the corresponding values for $b^{\varGamma \mu}_{hl \lambda}$ for each channel (cf. Eq.~\ref{eq:wavefunc}). We limited $l$ to a maximum value of $3$, resulting in values of $L$ up to $L=6$. The same information can be used to calculate the MF coeffients $C^{LM}_{P0}(\epsilon, \Delta q)$. The resulting MFPADs at $\epsilon = 7.6$~eV for $\theta = 0^\circ,45^\circ$ and $90^\circ$ are shown in Fig.~\ref{fig:MFPADs} (a),(b), and (c) respectively. (We note that adding an additional partial wave channel with $l = 5$ changes the calculated MFPADs by at most 3\%.)
 
  \begin{table}
    \centering
    \begin{tabular}{|c|c|c|c|c|c|}
    \hline
    Symmetry & $l$ & $\lambda$ & $b^{\varGamma \mu}_{hl\lambda}$ & $q$ & $D_{\zeta}$\\
         \hline \hline
         $\sigma_u$ & $1$ & $0$ & $1$ & $0$ & $1.9200 + 1.5840i$\\
         \hline 
         $\sigma_u$ & $3$ & $0$ & $1$ & $0$ & $-2.4750 + 1.0362i$\\
         \hline 
         $\pi_u$ & $1$ & $\mp1$ & $0.7071$ & $\pm1$ & $-1.1605 - 0.4746i$\\
         \hline 
         $\pi_u$ & $3$ & $\mp1$ & $0.7071$ & $\pm1$ & $1.5866 - 0.2253i$\\
         \hline 
    \end{tabular}
    \caption{Symmetrized matrix elements for $N_2$, a molecule with $D_{\infty h}$ symmetry, for the ionizing transition from the neutral ground state $N_2(X^{1}\Sigma^{+}_{g})$ to the ground ionic state $N^+_2(X^{2}\Sigma^{+}_{g})$. For this transition, the ejected electron can only be found in the $\sigma_u$ or $\pi_u$ state.}
    \label{tab:Dinfinitytable}
\end{table}
 
\begin{figure} 
     \centering
     \includegraphics[width=.80\textwidth,trim=6 6 6 6, clip]{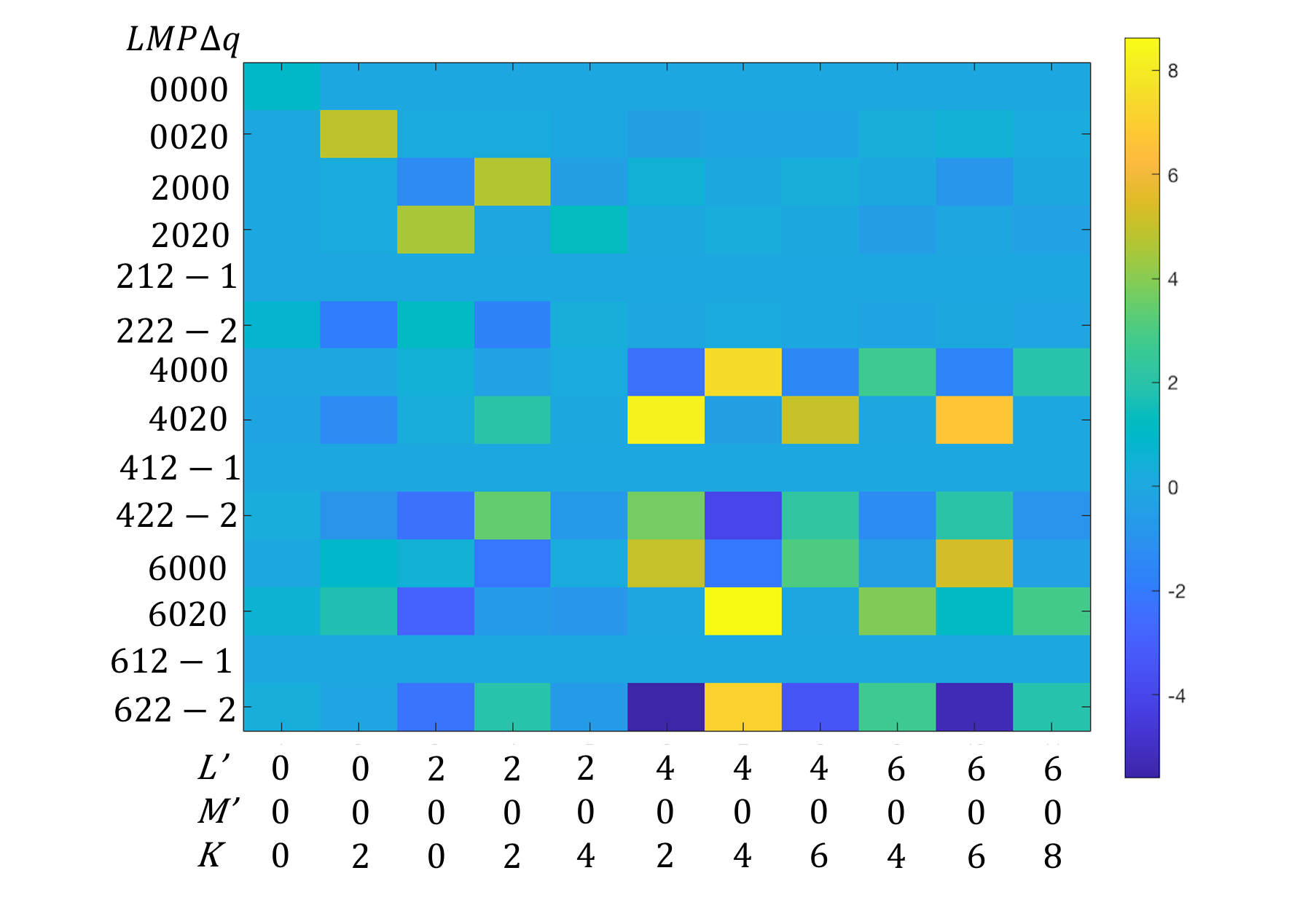}
     \caption{Values of $\mathbf{\hat{G}}^{LMP\Delta q }_{L'M'KS}$ matrix elements for $N_2$, for the ionizing transition from the ground state $N_2(X^{1}\Sigma^{+}_{g})$ to the ground ionic state $N^+_2(X^{2}\Sigma^{+}_{g})$. See text for details.}
     \label{fig:G}
 \end{figure}

 We then carried out the retrieval by  first calculating $\mathbf{\Gamma}^{\zeta\zeta'LM}_{P0\Delta q}$ and  $\mathbf{\Gamma}^{\zeta\zeta'L'0 }_{K00}$, and using them to build $\mathbf{\hat{G}}^{L0P\Delta q }_{L'0K0}$. Since $S = 0$ for a linear molecule, $\mathbf{\overline{C}}^{L' 0}_{K0} = \mathbf{C}^{L' 0}_{K0}$ in Eq.~\ref{eq:CfromG}. Figure~\ref{fig:G} shows the matrix $\mathbf{\hat{G}}^{LMP\Delta q }_{L'0K0}$ with corresponding ${LMP\Delta q}$ and ${L'M'K}$ ($M' = 0$ due to cylindrical symmetry as previously discussed) values, for allowed terms up to $l=3$. From the figure, it is seen that for $N_2$, rows corresponding to $M=-\Delta q=1$ in $\mathbf{\hat{G}}^{LMP\Delta q }_{L'0K0}$ are entirely zero. 
 This indicates that there are no equations linking these MF $C^{LM}_{P0}(\epsilon,\Delta q)$ to any LF $C^{L'0}_{K0}(\epsilon)$, rendering them irretrievable. Removing these leaves eleven $C^{L'0}_{K0}(\epsilon)$ and eleven $C^{LM}_{P0}(\epsilon,\Delta q)$, implying a consistent system of liner equations, Eq.~\ref{eq:CfromG}, with one independent linear equation for each unknown $C^{LM}_{P0}(\epsilon,\Delta q)$. It is therefore possible to retrieve these eleven $C^{LM}_{P0}(\epsilon,\Delta q)$ uniquely. With these we expect to only retrieve the MFPADs at $\theta = 0^\circ$ and $90^\circ$ since $\Delta q = 1$ coefficients do not contribute at either orientation due to the following properties of the Wigner D matrix elements: $D^P_{0\Delta q} (0,0,0) = \delta_{0\Delta q}$ and $D^P_{01} (0,\pi/2,0) = 0$. Within these constraints, Eq.~\ref{eq:CfromG}, highlighted in the red box in Fig.~\ref{fig:flowchart}, was used to retrieve the $C^{LM}_{P0}(\epsilon,\Delta q)$ staring with calculated LF coefficients $C^{L'0}_{K0}(\epsilon)$. Eq.~\ref{eq:betaMF} and Eq.~\ref{eq.MFPAD}, in the bottom two boxes in Fig.~\ref{fig:flowchart}, were then used to construct the retrieved MFPADs. Fig.~\ref{fig:MFPADs} (d),(e), and (f) show the resulting MFPADs for light polarized along the $z$-axis, at $\theta = 45^\circ$ with respect to the $z$ axes, and along the $x$ axis respectively (note that here any polarisation direction in the $(x,y)$ plane is equivalent).
 As is evident, we are only able to exactly reproduce the MFPADs for light polarized along the $z$ and $x$ axes from the LF $C^{L'0}_{K0}$, as expected. Physically, it is interesting to note that the missing information in this case corresponds to the phase difference (hence interference) between the $\sigma_u$ and $\pi_u$ continua, which is restricted due to the symmetry of the problem. The reconstruction in this case can actually be recognised as the symmetrized result in $D_{\infty h}$, in which cylindrical symmetry about the $z$-axis is imposed, and the result is averaged over all orientations of the $x$-axis. As previously mentioned, this is \emph{less} information than available from a complete experiment, which has been performed for this case in~\cite{marceau2017}, permitting \emph{all} MFPADs to be retrieved. However, no similar retrieval has ever been demonstrated for a polyatomic molecule, to which our methodology is easily extended, as shown by the results in the next section. It is also worth noting once more that similar ``information content" effects and issues are found in traditional complete experiments, in which certain terms may be irretrievable due to fundamental and/or experimentally-imposed symmetry restrictions~\cite{hockett2018QMP2}.

\subsection{Results for $D_{2h}$ Symmetry}

\begin{table}[H]
    \centering
    \begin{tabular}{|c|c|c|c|c|c|}
    \hline
    Symmetry & $l$ & $\lambda$ & $b^{\varGamma \mu}_{hl\lambda}$ & $q$ & $D_{\zeta}$\\
         \hline \hline
         $A_g$ & $0$ & $0$ & $1$ & $\mp1$ & $\pm(1.4827 - 1.8389i)$\\
         \hline 
         $A_g$ & $2$ & $0$ & $1$ & $\pm1$ & $\pm(1.7666 + 1.1582i)$\\
         \hline 
          $A_g$ & $2$ & $[-2,2]$ & $[0.7071,0.7071]$ & $\mp1$ & $\pm(2.6519 + 1.3927i)$\\
         \hline 
          $B_{1g}$ & $2$ & $[-2,2]$ & $[-0.7071,0.7071]$ & $\pm1$ & $2.1671 + 1.7278i$\\
         \hline 
          $B_{2g}$ & $2$ & $[-1,1]$ & $[-0.7071,0.7071]$ & $0$ & $-1.1444 + 0.0000i$\\
         \hline 
        
    \end{tabular}
    \caption{Symmetrized matrix elements for $C_2H_4$, a molecule with $D_{2h}$ symmetry, for ionizing transition from the ground state of $C_2H_4$ ($^1A_g$) to the ground state ($^2B_{3u}$) of $C_2H_4^+$. For this transition, the ejected electron can be found in the $A_g$, $B_{1g}$, or $B_{2g}$ states.}
    \label{tab:D2hsymmetry}
\end{table}

\begin{figure}
    \centering
    \includegraphics[width=1\textwidth,trim=15 15 15 15, clip]{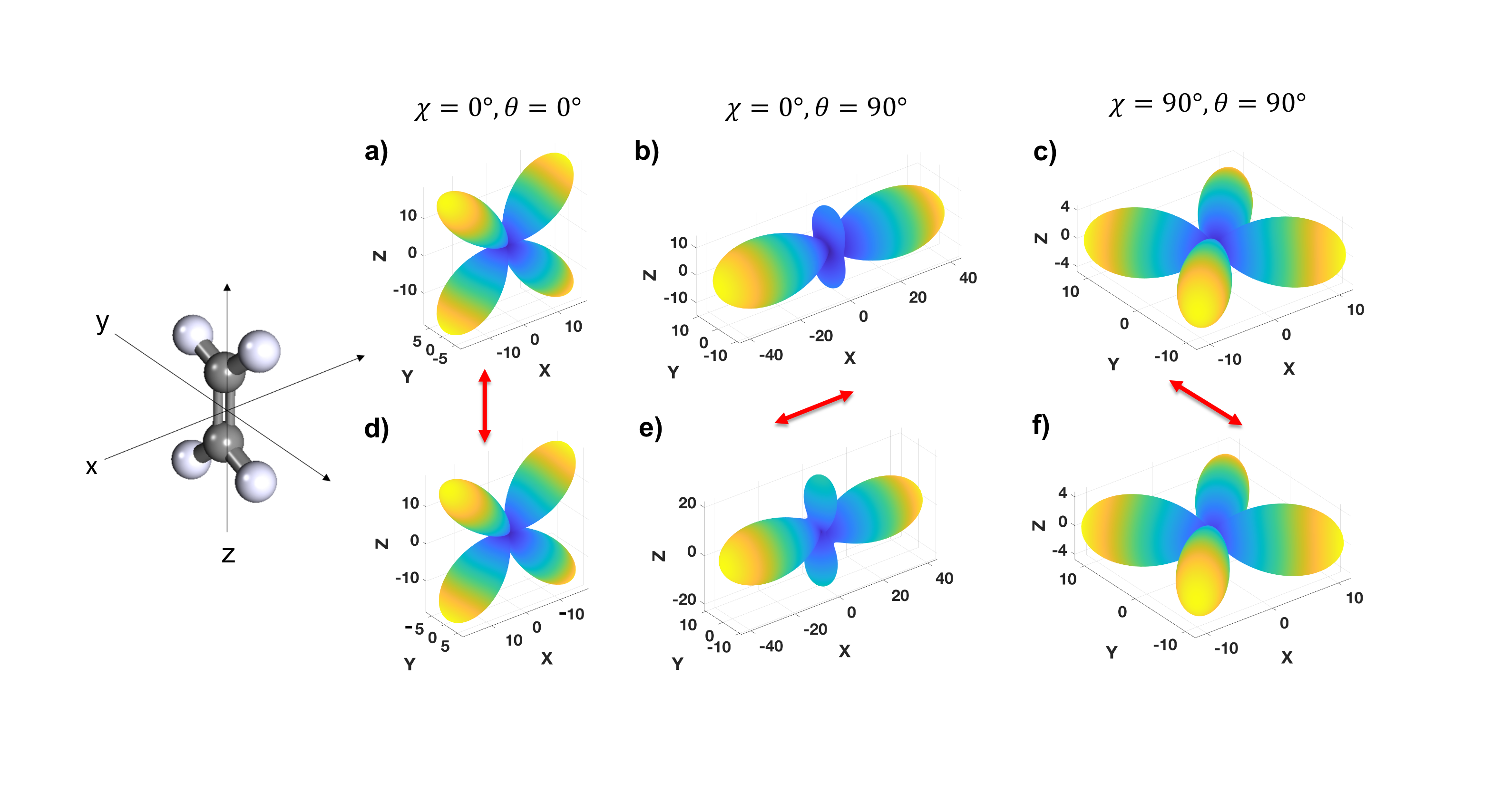}
    \caption{Calculated ((a),(b),(c)) and retrieved ((d),(e),(f)) MFPADs for $C_2 H_4$ for light polarized (red arrows) along the $z$, $x$, and $y$ axes, respectively. The MF is shown to the left, for the ground state molecular geometry. The $x$, $y$ and $z$ axes correspond to those in the MFPAD plots.}
    \label{fig:D2hMFPADs0}
\end{figure}

\begin{figure}
    \centering
    \includegraphics[width=.75\textwidth, trim=6 6 6 6, clip]{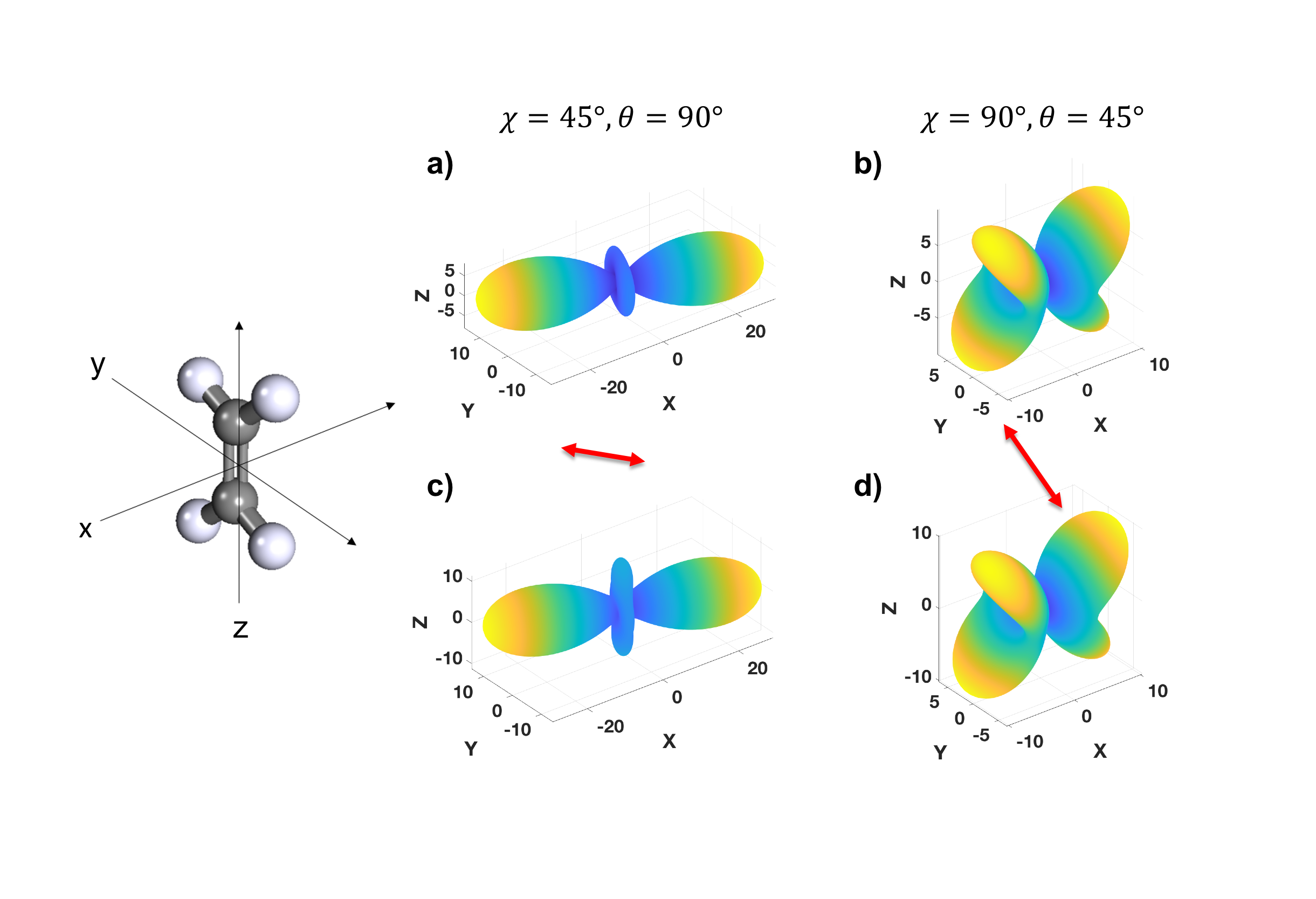}
    \caption{Calculated ((a),(b)) and retrieved ((c),(d)) MFPADs for $C_2 H_4$ for light polarized (red arrows) in the $xy$ plane and $zy$ plane, respectively. The MF is shown to the left, for the ground state molecular geometry. The $x$, $y$ and $z$ axes correspond to those in the MFPAD plots.}
    \label{fig:D2hMFPADs45}
\end{figure}

We choose $C_2H_4$ as an example of a molecule with $D_{2h}$ symmetry and demonstrate the retrieval of MFPADs for the $X^1A_g \rightarrow X^2B_{3u}$ ionization channel at a photoelectron energy of 1~eV. For $D_{2h}$, $S$ is even and the MFPADs are $\chi$ dependent, but still independent of the azimuthal angle $\phi$, as only linearly polarized light is considered, implying that $M'=0$ in the LF. For $C_2H_4$, we find by numerical calculation that $C^{L'0}_{KS}$=$C^{L'0}_{K-S}$ and as a result that $\overline{C}^{L'0}_{KS}=C^{L'0}_{KS}$. Calculations were carried out using a max $l$ value of $l=2$. Table~\ref{tab:D2hsymmetry} shows the possible symmetries of the ejected electron and the corresponding values of $b^{\varGamma \mu}_{h l \lambda}$. In this table, the corresponding values of $D_{\zeta}(\epsilon)$ were calculated using ePolyscat with $\epsilon =1$~eV. This results in eighteen $C^{L'0}_{KS}$ and twenty-seven $C^{LM}_{P0}(\epsilon,\Delta q)$, and no zero rows in the  $\mathbf{\hat{G}}^{LMP\Delta q }_{L'0KS}$. If the system of linear equations represented by Eq.~\ref{eq:CfromG} were consistent, the $C^{LM}_{P0}(\epsilon,\Delta q)$ would be uniquely retrievable. However, since there is a larger number of $C^{LM}_{P0}(\epsilon,\Delta q)$ than $C^{L'0}_{KS}$, we cannot uniquely determine the $C^{LM}_{P0}(\epsilon,\Delta q)$. Nonetheless, solving Eq.~\ref{eq:CfromG} by the Moore-Penrose inverse provides a least-squares fit of the $C^{L'0}_{KS}$ such that the retrieved vector $\mathbf{C}^{LM}_{P0}(\epsilon,\Delta q)$ has the smallest possible norm~\cite{planitz1979inconsistent}. This then still allows the construction of retrieved MFPADs which can be checked against MFPADs calculated directly from the ionization dipole matrix elements. In figures~\ref{fig:D2hMFPADs0} and \ref{fig:D2hMFPADs45} we compare the calculated MFPADs to the retrieved MFPADs for different polarization geometries in the MF. Figure ~\ref{fig:D2hMFPADs0} shows the MFPADs for light polarized along the $x$, $y$, and $z$ molecular axes. It is noted that for $\theta=0^\circ$, the MFPADs do not depend on the value of $\chi$. In Figure ~\ref{fig:D2hMFPADs45}, we plot the MFPADs for light polarized in the $xy$ and $zy$ planes. Though small differences are evident, in each case it is seen that the retrieved MFPADs closely resemble the calculated MFPADs of the same orientation.


The limit of $l= 2$ here, restricting $L$ to 4, is intended to limit $K$ to a potentially experimentally achievable maximum value of 6~\cite{lam2020,makhija2016}. The width of the rotational wavepacket excited by the pump pulse, quantified by its angular momentum $K$, therefore limits the resolution of the retrieved MFPAD. We note that adding an additional partial wave channel with $l =4$ does not qualitatively affect the shape of the MFPADs, the most significant effect being at the node of the $x$ polarized PAD.  A potential route to improving the retrieval would be to break symmetry in the LF by using an elliptically polarized pump pulse. Breaking cylindrical symmetry increases the number of $C^{L'M'}_{KQS}$, since $M'$ and $Q$ are no longer zero in the LF, while leaving the number of $C^{LM}_{P0}(\epsilon,\Delta q)$ unchanged in the MF. This could provide a better least-squares retrieval or potentially a consistent set of equations in Eq.~\ref{eq:CfromG}, resulting in a unique solution for $\mathbf{C}^{LM}_{P0}(\epsilon,\Delta q)$. However, the form of the $\Gamma^{\zeta\zeta'L'M'}_{K0S}$ and therefore $\mathbf{\hat{G}}^{LMP\Delta q }_{L'M'KS}$ must be re-derived for this case, with $Q\neq0$ and $M\neq0$. Again, it is of note that similar considerations and restrictions pertain, in general, to traditional complete experiment methodologies, and the precise details are also case-specific ~\cite{hockett2018QMP2}.

\section{Conclusion and Outlook}
\label{sec:Conclusion}
Time-resolved MFPADs are a powerful but challenging probe of valence electronic dynamics during ultrafast molecular processes.We described here a route to reconstructing MFPADs for a polyatomic molecule which avoids the burden of first determining the photoionization matrix elements. We employ, as raw material for the reconstruction, information from time resolved LF measurements of PADs from a RWP prepared in the ground electronic state of the molecule. Such a measurement provides the coefficients, the $\mathbf{C}^{lab}$ in Eq.~\ref{eq:basic}, a matrix equation which can potentially be used to retrieve the set of coefficients $\mathbf{C}^{mol}$, used to reconstruct the MFPADs. For inversion symmetric molecules containing a horizontal and vertical mirror plane ($D_{nh}$ point group symmetry) which are photoionized by linearly polarized light, we derived the matrix equation Eq.~\ref{eq:CfromG} - equivalent to Eq.~\ref{eq:basic} - shown in the central red box in Fig.~\ref{fig:flowchart} by systematic application of symmetry arguments. The coefficients $C^{L'0}_{KS}$ represent the measureable coefficients $C^{lab}$, and solving for the coefficients ${C}^{LM}_{P\Delta q}$ provides the MFPADs, importantly with no requirement for the `complete' determination of the complex dipole photoionization matrix elements. We numerically demonstrate this reconstruction procedure for two molecules $N_2$ and $C_2 H_4$, having point group symmetry $D_{\infty h}$ and $D_{2h}$, respectively. The results for $C_2 H_4$ in Figs~\ref{fig:D2hMFPADs0} and \ref{fig:D2hMFPADs45} show that this method can faithfully reconstruct the MFPADs for a polyatomic molecule. In a future direction, we will examine the sensitivity of our MFPADs reconstruction procedure to noise in the experimental data, an important issue when dealing with real experimental data. To this end, we note that several numerical methods are available to obtain a least-squares solution to Eq.~\ref{eq:CfromG} (red box in Fig.~\ref{fig:flowchart}) even for an inconsistent system of equations~\cite{boyd2011,li2018,driver2020}.

We expect this method to be particularly useful for polyatomic molecules. Recent studies have shown that a molecular ensemble needs to be nearly perfectly aligned before a significant difference can be observed between PADs from aligned and isotropic ensembles~\cite{reid2018}. This, in general, is very difficult to achieve under field free conditions, particularly for all three axes of a polyatomic molecule. In contrast, the method presented here does not necessarily demand a high degree of alignment, but rather requires only that large enough values of $K$ be excited - up to $K=8$ for $N_2$ and $K = 6$ for $C_2 H_4$ for the reconstructions presented here - corresponding to a broad wavepacket of rotational states. The degree of alignment is then determined by the time evolution of the relative phases of these states under field free conditions. Since linear molecules have harmonically spaced rotational energy levels, high degrees of alignment automatically ensue and repeat periodically. The rotational level spacings for polyatomic molecules, on the other hand, are not necessarily harmonic, leading to complex, aperiodic rotational dynamics from a broad rotational wavepacket, not necessarily providing a high degree of alignment. Nevertheless, recent experimental studies indicate that the $\mathbf{C}^{lab}$ coefficients can be faithfully retrieved from such a wavepacket, given sufficient signal-to-noise in the data~\cite{lam2020,makhija2016}. Therefore, we conclude that the fundamental requirements for the experimental realization of our MFPADs reconstruction method are: (i) a low initial rotational temperature, (ii) a pump-pulse (or pulse sequence~\cite{marceau2017,ren2014,ren2013,leibscher2004enhanced,bisgaard2004observation,cryan2009,cryan2010auger,pabst2010}) which produces a broad rotational wavepacket, and (iii) high fidelity data. We have also noted that a potential route to improving our reconstruction method would be to break cylindrical symmetry in the LF through use of an elliptically polarized pump pulse or time-separated cross-polarized `kick' pulses. This would preclude the use of standard photoelectron imaging techniques such as (2D) VMI which commonly use algorithms requiring cylindrical symmetry. However, we note that a variety of 3D VMI approaches have now been developed~\cite{Wollenhaupt2009,lee2014,debrah2020}, also in development in our lab, which circumvent this experimental symmetry restriction. 

We hope that these results will benefit the study of photoionization dynamics of ground state polyatomic molecules - particularly in regions of the continuum containing resonances, which are particularly sensitive to molecular orientation~\cite{shigemasa1995,bellm2005,bertrand2012,staniforth2013,ren2013, kamalov2020}. We also anticipate that this approach will benefit the study of excited state dynamics. In both cases MFPADs are a valuable source of information, and the results presented here allow this information to be readily accessed.      
\\
\\
\emph{\textbf{Acknowledgments}}\\
VM and PH thank Vinod Kumarappan for valuable discussions. VM and MG acknowledge the Summer Science Institute at University of Mary Washington for funding. AS thanks the NSERC Discovery Grants program and the NRC-uOttawa JCEP for financial support.
\\
\\
\section{Appendix}\label{Sec:Appendix}
\subsection{Derivation of Permutation identities for the Coupling Parameters}
The $\Gamma^{\zeta\zeta'LM}_{K0S}$ parameter, first seen in Eq.~\ref{eq:coeffLF} in section~\ref{sec:AF-ionization}, is expanded below: 
\begin{equation}
\begin{split}
    \Gamma^{\zeta\zeta'LM}_{K0S}=\sum_P\sqrt{2L+1}(-1)^{K}(-1)^P(2P+1)\begin{pmatrix}1&1&P\\0&0&0 \end{pmatrix}\begin{pmatrix}P&K&L\\M&0&M \end{pmatrix}(-1)^q
    \\
    \begin{pmatrix}1&1&P\\q&-q'&q'-q \end{pmatrix}
\begin{pmatrix}P&K&L\\q'-q&-S&S+q-q' \end{pmatrix}\sqrt{(2l+1)(2l'+1)}\begin{pmatrix}l&l'&L\\0&0&0 \end{pmatrix} B^{\varGamma \mu hl}_{\varGamma' \mu' h'l'}(L,S,q,q'),
\label{eq:FullLFGamma}
\end{split}
\end{equation}
In Eq.~\ref{eq:FullLFGamma}, the term $B^{\varGamma \mu hl}_{\varGamma' \mu' h'l'}(L,S,q,q')$ contains the sum over $\lambda$ and $\lambda'$. 
\begin{equation}
    B^{\varGamma \mu lh}_{\varGamma' \mu' l'h'}(L,S, q, q') =\sum_{\lambda\lambda'} (-1)^{\lambda'}\begin{pmatrix}l&l'&L\\-\lambda&\lambda'&S+q-q'\end{pmatrix}b^{\varGamma\mu}_{hl\lambda}b^{\varGamma'\mu' *}_{h'l'\lambda'}.
    \label{eq:BLF}
\end{equation}
 It is from the Wigner 3j symbols contained in Eqs.~\ref{eq:FullLFGamma} and~\ref{eq:BLF} that the LF selection rules presented in Section ~\ref{sec:AF-ionization} are determined. Using the behavior of the Wigner 3j symbol in Eq.~\ref{eq:BLF} under odd permutation of the columns, time reversal, and that its second row sums to zero, the following identity can be determined:
 \begin{equation}
    B^{\varGamma' \mu' l'h'}_{\varGamma \mu lh}(L,S, q',q) =(-1)^{-S-q'+q}B^{\varGamma \mu lh*}_{\varGamma' \mu' l'h'}(L,-S, q,q')
    \label{eq:swapBLF}
\end{equation}
This and the permutation identities for the 3j symbols lead to the general relation given in Eq.~\ref{eq: LFgeneral} in section~\ref{sec:AF-MF-theory}.
 \\
 \\
The parameter $\Gamma^{\zeta\zeta'LM}_{P\Delta q}$ introduced in Eq.~\ref{eq:coeffLF} can be written out as,
\begin{equation}
\begin{split}
    \Gamma^{\zeta\zeta'LM}_{P\Delta q}=\sqrt{2L+1}(2P+1)(-1)^{q'}\begin{pmatrix}1&1&P\\q&-q'&\Delta q \end{pmatrix}\begin{pmatrix}1&1&P\\0&0&0 \end{pmatrix}
    \\
    \sqrt{(2l+1)(2l'+1)}\begin{pmatrix}l&l'&L\\0&0&0 \end{pmatrix} B^{\varGamma \mu lh}_{\varGamma' \mu' l'h'}(L,M),
    \label{eq:MFFullGamma}
    \end{split}
\end{equation}
In this case, the term $B^{\varGamma \mu lh}_{\varGamma' \mu' l'h'}(L,M)$ contains the sum over $\lambda$ and $\lambda'$:
\begin{equation}
     B^{\varGamma \mu lh}_{\varGamma' \mu' l'h'}(L,M) =\sum_{\lambda\lambda'}(-1)^{\lambda}\begin{pmatrix}l&l'&L\\\lambda&-\lambda'&M\end{pmatrix}b^{\varGamma\mu}_{hl\lambda}b^{\varGamma'\mu' *}_{h'l'\lambda'}.
     \label{eq:BMF}
\end{equation}
The MF selection rules discussed is Section~\ref{sec:MF-ionization} originate from the Wigner 3j symbols given in Eqs.~\ref{eq:MFFullGamma} and~\ref{eq:BMF}.
Like with Eq.~\ref{eq:BLF}, we use Eq.~\ref{eq:BMF}, the behavior of the Wigner 3j symbol under odd permutation of the columns, time reversal, and that its second row sums to zero to determine:
\begin{equation}
    B^{\varGamma' \mu' l'h'}_{\varGamma \mu lh}(L,M) =(-1)^{M}B^{\varGamma \mu lh*}_{\varGamma' \mu' l'h'}(L,-M)
    \label{eq:swapBMF}
\end{equation}

Again, this and  and the permutation identities for the 3j symbols lead to the general relation given in Eq.~\ref{eq: GeneralMF}.

\subsection{Molecular Frame Symmetry Arguments}

In the MF, for all inversion symmetric molecules with a horizontal mirror plane,
\begin{equation}
\sigma(\epsilon, \theta'_{e},\phi'_{e}; \phi , \theta, \chi)=\sigma(\epsilon, \pi - \theta'_{e},\pi-\phi'_{e}; \phi , \theta, \chi)
\label{eq:Psymmetry}
\end{equation}
This symmetry, with the relation that $Y_{LM}(\pi - \theta'_{e},\pi - \phi'_{e})=(-1)^{L-M}Y_{L-M}(\theta'_e, \phi'_e)$, leads to the following relation between the  $\beta_{LM}$'s,

\begin{equation}
\beta_{LM}(\epsilon, \phi, \theta, \chi)=(-1)^{L-M}\beta_{L-M}(\epsilon, \phi, \theta, \chi).
\label{eq:Bequal2}
\end{equation}
Substituting this, and the relation and the relation  $Y_{L-M} = (-1)^M Y_{LM}^*$ in Eq.~\ref{eq: GeneralMF}, leads to the further determination that the  $\beta_{LM}(\epsilon, \phi, \theta, \chi)$ must be real since $\sigma(\epsilon, \theta'_{e},\phi'_{e}; \phi , \theta, \chi)$ is necessarily real.

Further, for molecules with inversion symmetry, inversion is equivalent to the rotation $\phi \rightarrow \phi + \pi$, $\theta \rightarrow \pi - \theta$ and $\chi \rightarrow \pi - \chi$ of an arbitrarily oriented molecule, followed by reflection of the through the molecular plane. The $\beta_{LM}(\epsilon, \phi, \theta, \chi)$ must remain invariant under this rotation. The Wigner rotation matrix elements in Eq.~\ref{eq:betaMF} behave as $D^{P}_{R \Delta q}(\phi + \pi,\pi - \theta,\pi - \chi) = (-1)^{P-\Delta q} D^{P}_{R -\Delta q}(\phi,\theta,\chi)$, leading to the following equation for the $C^{LM}_{PR}(\epsilon,\Delta q)$  
\begin{equation}
C^{LM}_{PR}(\epsilon,\Delta q)=(-1)^{P + M - \Delta q}C^{L-M}_{PR}(\epsilon,-\Delta q),
\label{eq:Cequal}
\end{equation}
after taking Eq.~\ref{eq:Bequal2} into account. For linearly polarized ionizing light, $P$ = 0 or 2, and $R=0$ lead to the identity 
\begin{equation}
C^{LM}_{P0}(\epsilon,\Delta q)=(-1)^{M - \Delta q}C^{L-M}_{P0}(\epsilon,-\Delta q).
\end{equation}
As given in Eq.~\ref{eq:Cequallin}.
\\
\\
\emph{\textbf{Data Availability Statement}}\\
The data that support the findings of this study are available from the corresponding author upon reasonable request.

\selectlanguage{english}
\FloatBarrier
\bibliographystyle{unsrt}  

\bibliography{references_171120_PH-250121}  
\end{document}